\documentstyle[12pt,titlepage]{article}
\begin{document}
\begin{titlepage}
\parindent 0pt
{\large\bf
Finite-size effects of dimensional crossover
in quasi-two-dimensional three-state Potts model
}

\bigskip

{\large\bf Atsushi Yamagata}

{\it
Department of Physics, Tokyo Institute of Technology,
Oh-okayama, Meguro-ku, Tokyo 152, Japan
}

\bigskip

\begin{description}
\item[Running title] Quasi-two-dimensional three-state Potts model

\item[Keywords] Quasi-two-dimensional magnet,
Three-state Potts model, Finite-size scaling,
Dimensional crossover, Monte Carlo simulation

\item[PACS classification codes] 02.70.Lq, 64.60.Cn, 75.10.Hk
\end{description}

\bigskip

{\bf Abstract}

A nearest neighbour spin pair
of the quasi-two-dimensional three-state Potts model
interacts with the strength $J(>0)$ in the $xy$-plane and
with $\lambda J$ $(0\le \lambda \ll 1)$ in the $z$-axis.
The phase transition is of second-order when $\lambda = 0$
and is of first-order when $\lambda > 0$.
The dimensional crossover occurs with a change of the order of
the phase transition.
We study the finite-size effects of the phenomenon
by using a Monte Carlo method with a multi-spin coding technique.
The prediction of the finite-size scaling theory
is consistent with the Monte Carlo results.
\end{titlepage}

\section{Introduction}
A quasi-two-dimensional system is a three-dimensional one
in which the ratio $\lambda$ of the interplanar to
the intraplanar exchange interactions is small.
If the system exhibits a phase transition,
we can see a crossover from two-di\-mensional to
three-dimensional behaviour as the critical point is approached.
Many authors have studied quasi-two-dimensional antiferromagnets
by neutron scattering experiments
\cite{Collins89}.

The theoretical studies of the dimensional crossover
have been carried out by using
perturbation theory
\cite{Nesis65}-\cite{Coniglio72},
high temperature series expansion
\cite{PaulStanley71}-\cite{Krasnow73},
generalized homogeneous functions
\cite{HankeyStanley72},
and
rigorous approach
\cite{LiuStanley72}-\cite{CitteurKasteleyn72}.
The results are that the critical and the crossover temperature
are singular with respect to $\lambda$.
They behave as $\lambda^{1/\phi}$ where $\phi$ is
the crossover exponent
\cite{RiedelWegner69}.
It has been shown that $\phi$ is equal to
the critical exponent $\gamma$ for the susceptibility of
the system with $\lambda=0$.
The present author discussed finite-size scaling and
performed Monte Carlo simulaltions of
the quasi-two-dimensional Ising model for the first time
\cite{Yamagata94}.

In the above mentioned researches
there was an assumption that the order of the phase transition
was unaltered.
In this paper
we study finite-size scaling of the dimensional crossover
in which it changes.
We perform Monte Carlo simulations
of the three-state ferromagnetic Potts model
\cite{Wu82}.
The phase transition is of second-order in two dimensions
\cite{Baxter73}
and is of first-order in three dimensions
\cite{Yamagata93}.
In the next section we review the finite-size scaling theory
for the quasi-two-dimensional systems briefly.
The finite-size scaling form of an effective transition temperature
is presented.
In section \ref{sec:mcs} we describe an algorithm
for the multi-spin coding technique used
in our Monte Carlo simulations
of the quasi-two-dimensional three-state Potts model.
The Monte Carlo data are compared with the prediction of
the finite-size scaling theory in section~\ref{sec:mcr}.
A summary is given in section \ref{sec:sum}.

\section{Finite-size scaling theory}
\label{sec:fsst}
We review the finite-size scaling theory
for the quasi-two-dimensional systems briefly.
The detailed discussion is in the reference
\cite{Yamagata94}.
Let us consider the three-state Potts model
on the simple cubic lattice for the sake of concreteness.
The Hamiltonian is
\begin{equation}
{\cal H}_{\lambda}
=
\sum_{\langle ij \rangle}J_{ij}\,[1-\delta(\sigma_{i},\sigma_{j})]
+H \sum_{i} [1-\delta(\sigma_{i},1)]
\label{eqn:hamilton}
\end{equation}
where $\sigma_{i}$ is a Potts spin variable located
{\em i\/}th lattice site
and which takes on the value 1, 2, and 3.
The first summation is over all nearest neighbour pairs
on the lattice, the second summation over all lattice sites.
The strength $J_{ij}$ of the interaction
for the nearest neighbour pair $ij$ is $J(>0)$ in the $xy$-plane and
$\lambda J$ $(0\le \lambda \ll 1)$ in the $z$-axis.
$H$ is an external magnetic field.
When $\lambda=0$,
the equation (\ref{eqn:hamilton}) consists of
two-dimensional three-state ferromagnetic Potts models
which are independent of each other.

We assume that the free energy per spin measured by $k_{\rm B}T$,
where $k_{\rm B}$ is Boltzmann's constant and $T$ is the temperature,
is a generalized homogeneous function
of variables
\[
t_{0}=T/T(0)-1,
\]
$h=H/k_{\rm B}T$, and $\lambda$ as $t_{0}$, $h$, $\lambda\to 0$,
where $T(0)$ is the critical temperature of the system
with $\lambda=0$.
We can derive
\begin{equation}
f(t_{0},h,\lambda)=|t_{0}|^{2-\alpha}\,
f^{\pm}(h/|t_{0}|^{\beta+\gamma},\lambda/|t_{0}|^{\phi})
\label{eqn:scalethlmd}
\end{equation}
with a scaling function $f^{\pm}(x,y)$
where $+$ ($-$) refers to $t_{0}>0$ ($t_{0}<0$),
and $\alpha+2\beta+\gamma=2$ where $\alpha$, $\beta$, and $\gamma$
are the critical exponents of the two-dimensional system
for the specific heat, the magnetization, and the susceptibility,
respectively.
The number $\phi$ is the crossover exponent and
is equal to $\gamma$. From (\ref{eqn:scalethlmd})
we get the behaviour of $T(\lambda)$
that is the transition temperature of the system with $\lambda$
as follows.
\begin{equation}
T(\lambda)/T(0)-1=A^{T}\,\lambda^{1/\phi}
\label{eqn:tlmd}
\end{equation}
where $A^{T}$ is a constant.

Let us consider the three-state Potts model (\ref{eqn:hamilton})
on an $L\times L\times L$ simple cubic lattice
to see the finite-size effects
\cite{Fisher70}-\cite{Privman90}
of (\ref{eqn:tlmd}).
To avoid surface effects we impose periodic boundary conditions.
We assume that
the free energy per spin is a generalized homogeneous function
of $t_{0}$, $h$, $\lambda$, and $L$
as $t_{0}$, $h$, $\lambda$, $1/L\to 0$ and
the system is characterized by $L/\xi(t_{0})$
where $\xi(t_{0})$ is the correlation length of
the system of $h=\lambda=1/L=0$.
Using (\ref{eqn:scalethlmd})
we can derive
\begin{equation}
f(t_{0},h,\lambda,L)
=
L^{-(2-\alpha)/\nu}\,
\tilde{f}(t_{0}\,L^{1/\nu},h\,
L^{(\beta+\gamma)/\nu},\lambda\,L^{\phi/\nu})
\label{eqn:fssl}
\end{equation}
where $\tilde{f}(x,y,z)$ is a scaling function and
$\nu$ is the critical exponent for $\xi(t_{0})$.

Let us define an effective transition temperature as the position,
$T_{L}(\lambda)$, of the peak of the specific heat. From
(\ref{eqn:fssl}) we get
\begin{equation}
T_{L}(\lambda)/T(0)-1=L^{-1/\nu}\,\tilde{T}(\lambda\,L^{\phi/\nu})
\label{eqn:tlmdl}
\end{equation}
where $\tilde{T}(x)$ is a scaling function.
If $\tilde{T}(x)\to A^{T}\,x^{1/\phi}$ as $x\to +\infty$,
the equation (\ref{eqn:tlmd}) is reproduced
in the limit $L \to +\infty$
for a fixed value of $\lambda(>0)$.
The prediction (\ref{eqn:tlmdl}) will be compared with
the Monte Carlo data in section \ref{sec:mcr}.

\section{Monte Carlo simulations}
\label{sec:mcs}
To confirm the prediction (\ref{eqn:tlmdl})
of the finite-size scaling theory,
we perform Monte Carlo simulations
\cite{Binder79,BinderStauffer87}
of the quasi-two-dimensional three-state Potts model
(\ref{eqn:hamilton}) in $H=0$
on the $L \times L \times L$ simple cubic lattice
under fully periodic boundary conditions.
We use a multi-spin coding technique
\cite{Bhanot86a,Bhanot86b}
to simulate a large number of systems simultaneously.
Since the FORTRAN compiler on the HITAC S-820/80 computer,
we have used, treats 32-bit integers,
we can update 32 systems independently.
Three-state Potts spin variables located at identical lattice sites
are stored in the 32-bit positions of two words
\cite{KikuchiOkabe92,OkabeKikuchi92}.

An algorithm is as follows.
Let us consider a flip of a spin $\sigma_{0}$.
The change in the energy on flipping the spin,
$\sigma_{0}\to \sigma'_{0}$, is
\begin{eqnarray*}
{\mit \Delta}E/J
& = &
-\sum_{j=1}^{4}
\{
 [1-\delta (\sigma_{0},\sigma_{j})]
-[1-\delta (\sigma'_{0},\sigma_{j})]
\}\\
& &
-\lambda \sum_{k=5}^{6}
\{
 [1-\delta (\sigma_{0},\sigma_{k})]
-[1-\delta (\sigma'_{0},\sigma_{k})]
\}
\end{eqnarray*}
where $\sigma_{j}$ ($j=1,2,3,4$) is
the nearest neighbour spin of $\sigma_{0}$
in the $xy$-plane and $\sigma_{k}$ ($k=5,6$) in the $z$-axis.
Using variables
\[
n_{xy} \equiv \sum_{j=1}^{4}
\{
 [1-\delta (\sigma_{0},\sigma_{j})]
-[1-\delta (\sigma'_{0},\sigma_{j})] +1
\}
\]
and
\[
n_{z} \equiv \sum_{k=5}^{6}
\{
 [1-\delta (\sigma_{0},\sigma_{k})]
-[1-\delta (\sigma'_{0},\sigma_{k})] +1
\},
\]
we have
\[
-{\mit \Delta}E/J = n_{xy}+\lambda\,n_{z}-4-2\lambda.
\]
Defining a variable $n\equiv 5n_{xy}+n_{z}+10$,
we may see the energy change as a function of $n$:
$(-{\mit \Delta}E/J)_{n}, n\in \{10,11,\ldots,54 \}$.

Using a random variable $W$ which takes an integer value,
we flip the spin if $n+w\ge 32$ where $w$ is a possible value of
$W$.
The distribution of $W$ is determined by
\begin{eqnarray*}
\mbox{Prob}\{W\ge 32-n\}
&=&
\min(e^{K(-{\mit \Delta}E/J)_{n}},1)\\
&=&
\left\{
\begin{array}{ll}
e^{K(-{\mit \Delta}E/J)_{n}},&10\le n<32,\\
1,&32\le n\le 54,
\end{array}
\right.
\end{eqnarray*}
where $K=J/k_{\rm B}T$.
Thus the update is accepted with the probability
$\min(e^{-\beta{\mit \Delta}E},1)$ where $\beta=1/k_{\rm B}T$.
This is the same procedure as of the Metropolis algorithm
\cite{Metropolis53}.
We express $n+w$ $(\in \{ 10,11,\ldots,76 \})$,
which can be calculated with logical operations,
as the binary notation:
$\sum_{l=0}^{6}x_{l}2^{l}$, $x_{l}\in\{0, 1\}$.
In our algorithm we carry out the update when $x_{5}=1$ or $x_{6}=1$
although the Metropolis procedure needs to refer to the inequality
$r\le \min(e^{-\beta{\mit \Delta}E},1)$ where $r$ is
a possible value of
a random variable $R$ with uniform distribution over [0,1].
The algorithm explained here is for $\lambda \in (0,1/4)$.
It is sufficient to study the system since we are interested in
the case of the small value of $\lambda$.

The pseudorandom numbers are generated by
the Tausworthe method
\cite{ItoKanada88,ItoKanada90}.
We measure physical quantities at a temperature
over $10^{5}$ Monte Carlo steps per spin
(MCS/spin) after discarding $10^{4}$ MCS/spin to attain equilibrium.
Physical quantities are calculated by
the multi-step bitwise summation algorithm
\cite{ItoKanada88,ItoKanada90}.
Our algorithm achieves a speed of 25 million spins per second
on a $30\times 30\times 30$ lattice with measurements at every step.
Let us denote the average of a physical quantity, $O$,
in each system by $\langle O \rangle_{i}$, $i=1,2,\ldots,32$.
The expectation value is given by
\[
\overline{\langle O \rangle}
=
\frac{1}{32} \sum_{i=1}^{32} \langle O \rangle_{i},
\]
the standard deviation by

\[
{\mit \Delta}\langle O \rangle
=\left( \overline{\langle O \rangle^{2}}
-\overline{\langle O \rangle}^{2} \right)^{1/2}/\sqrt{31}.
\]

\section{Monte Carlo results}
\label{sec:mcr}
In the reference
\cite{Yamagata94}
we reported that there was hysteresis in magnetic quantities
(magnetization, susceptibility, $\ldots$ )
but we could not see it in the energy,
the specific heat:
\[
C=k_{\rm B}\,\beta^{2}\,
\left(
\overline{\langle {\cal H}_{\lambda}^{2}\rangle}
-\overline{\langle {\cal H}_{\lambda}\rangle^{2}}
\right)
/L^{3},
\]
and the fourth-order cumulant of the energy
for the small value of $\lambda$ or a small number of MCS/spin.
The hysteresis vanished for the large value of $\lambda$ or
a large number of MCS/spin.
We have observed similar behaviour for the quasi-two-dimensional
three-state Potts model.
According to
\cite{Yamagata94},
we analyse the peak position, $T_{L}(\lambda)$,
of the specific heat
since we can obtain useful information
about phase transitions from it.

Figure 1 shows the temperature dependence of $C$ of $L=20$ system
for various $\lambda$.
The solid curves are obtained by the smoothing procedure of
the fourth-order {\em B}-spline
\cite{Tsuda88}.
As $\lambda$ increases,
the shape of the curve becomes sharper and
the peak position $T_{20}(\lambda)$ becomes higher.
Then we get the peak height and $T_{20}(\lambda)$.

Figure 2 shows the finite-size scaling plot of $T_{L}(\lambda)$.
We have used the exact values of the critical temperature,
$k_{\rm B}T(0)/J=1/\ln (1+\sqrt{3})$,
and the critical exponents,
$\phi=\gamma=13/9$ and $\nu=5/6$,
of the two-dimensional three-state ferromagnetic Potts model
\cite{Wu82}.
The data fall on a common curve.
It is consistent with (\ref{eqn:tlmdl}).
The slope seems to approach $1/\phi=1/\gamma=9/13$.
It indicates that
the asymptotic behaviour of the finite-size scaling function
is $\tilde{T}(x)\to \mbox{const.} \times x^{1/\phi}$
as $x\to +\infty$.

\section{Summary}
\label{sec:sum}
The dimensional crossover occurs with a change
of the order of the phase transition
in the quasi-two-dimensional three-state Potts model.
We have studied the finite-size effects of the phenomenon
by using a Monte Carlo method.
An algorithm of a multi-spin coding technique
for this model has been presented.
The prediction (\ref{eqn:tlmdl}) of the finite-size scaling theory
has been consistent with the Monte Carlo results.
We have seen the asymptotic behaviour
of the finite-size scaling function
on the system with $\lambda \le 0.24 $ and $L \le 30$.

\section*{Acknowledgements}
The author would like to thank Dr. Katsumi Kasono
for useful discussions and critical reading of the manuscript.
We performed the simulations on the HITAC S-820/80 computer
of the Computer Centre, Institute for Molecular Science,
Okazaki National Research Institutes and of the Computer Centre
at Hokkaido University.
This study was supported by a Grant-in-Aid
for Scientific Research from the Ministry of Education, Science,
and Culture, Japan.


\clearpage
\section*{Figure captions}
\begin{description}
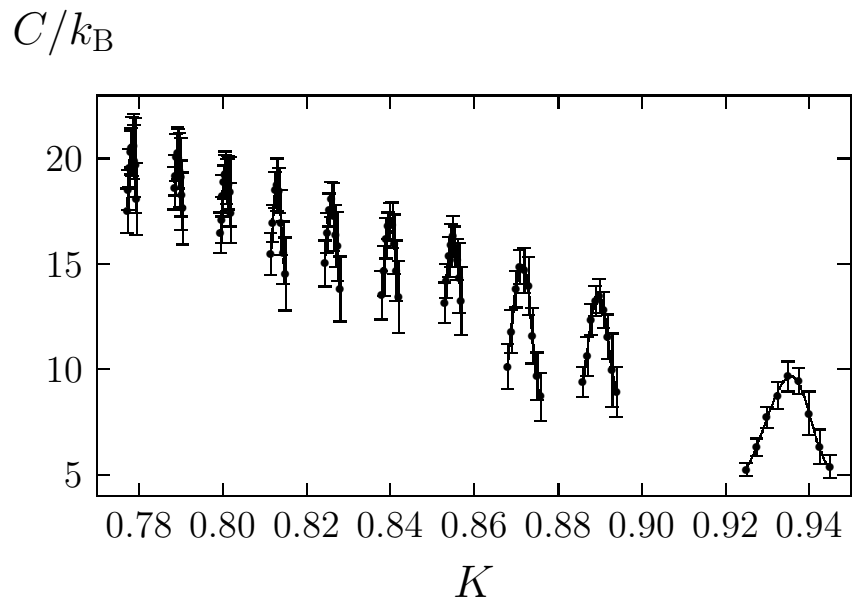
\item[Figure 1] Temperature ($K=J/k_{\rm B}T$) dependence of
the specific heat of
the $L=20$ system for various $\lambda$:
0.04,
0.08,
0.1,
0.12,
0.14,
0.16,
0.18,
0.20,
0.22,
0.24.
The solid curves are obtained by the smoothing procedure of the
fourth-order {\em B}-spline.
As $\lambda$ increases,
the shape of the curve is sharper and
$T_{20}(\lambda)$ becomes higher.

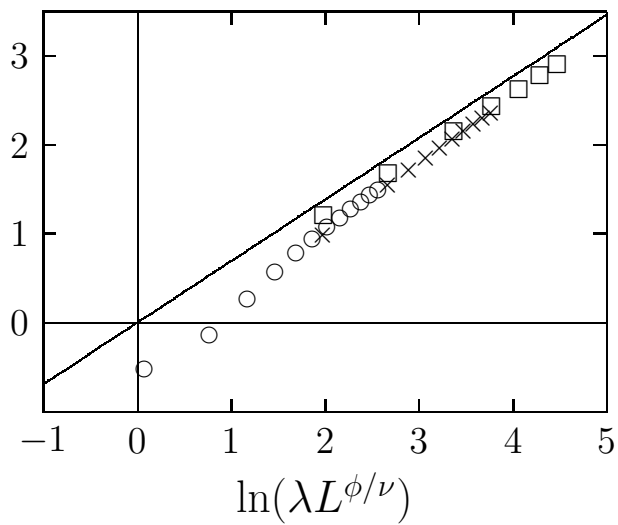
\item[Figure 2] Finite-size scaling plot of
the effective transition temperature.
The data of the system with $L=10$, 20, and 30 are denoted by
$\bigcirc$, $\times$, and $\Box$, respectively.
We have set that $k_{\rm B}T(0)/J=1/\ln (1+\sqrt{3})$,
$\phi=\gamma=13/9$, and $\nu=5/6$.
The data fall on a common curve.
The solid line shows that $(9/13)\,\ln (\lambda\,L^{26/15})$.
Errors are less than the symbol size.
\end{description}

\clearpage
\begin{figure}
\begin{center}
\setlength{\unitlength}{0.240900pt}
\ifx\plotpoint\undefined\newsavebox{\plotpoint}\fi
\sbox{\plotpoint}{\rule[-0.200pt]{0.400pt}{0.400pt}}%

\end{center}
\end{figure}
\end{document}